# Internet of Things (IoT) based Smart Agriculture Aiming to Achieve Sustainable Goals

**Dewan Md Nur Anjum Ashir**

Software Engineer

**Dr. Md. Taimur Ahad**

Assistant Professor

Computer Science Department

American International University (AIUB), Bangladesh

**Manosh Talukder**

American International University (AIUB), Bangladesh

**Tahsinur Rahman**

American International University (AIUB), Bangladesh

*Abstract— Despite the fact, a handful of scholars have endorsed the Internet of Things (IoT) as an effective transformative tool for shifting traditional farming to smart farming, relatively little study has addressed the enabling role of smart agriculture in achieving sustainable agriculture and green climate. Researchers are more focused on technological invention and model introduction rather than discussing societal or global development goals. Sustainable development goals (SDGs) designed by United Nations (UN), therefore demand discussions as SDGs targets have a closer implication of technology. To fill this gap, in this study a model of smart agriculture is developed and centring the model we investigated how the model addresses SDGs targets. The investigation suggests that smart agriculture supports targets mentioned in Goal 6, 7, 8, 9, 11 and 12 of SDG. This research is very important, both for developing and developed nations since most of the nations are moving more towards industrialization and aiming to achieve the SDG goals This research is expected to provide a path to the IT practitioners, governments and developing agencies on how technological intervention can provide a more sustainable agricultural world.*

*Keywords: IoT, Smart Agriculture, SDG. SAS, WSN, RFID, Zigbee 5.8, SDGs, United Nations (UN).*

## I. INTRODUCTION

The Sustainable Development Goals (SDGs) are a set of inter-multi-trans dimensional collections of seventy blueprint goals to achieve a better and sustainable world.The SDG aims to create a sustainable world by eradicating global poverty, safeguarding the environment and nature, and promoting global peace and prosperity. The observable detrimental social and global concerns such as food scarcity, high industrialization and low concerns regarding the climate impact, forwarded for such goals to induce.

A closer look into the targets declared in SDGs suggest that technology has a direct or indirect link to the SDGs. Earlier Kates et al. [1] also purported the envision of technology in SDG as a driver for sustainable community development. To enhance the speed for achieving the SDGs goals conceptual and theoretical framework models shouldbe devised. Technology implication thus becomes a vital factor in the realization of the SDGs.

Following the concept of technology implication for digitalization, agriculture hasbeen a topic in research, a transformative approach has been implemented in agriculture - from primitive agriculture to a smart agriculture system. Smart agriculture is a farming concept of IT-enabled infrastructure to leverage advanced technology – including the internet of things (IoT), big data and cloud computing for tracking, monitoring, automating and analyzing operations. Following the theme of smart agriculture, researchers such as [2][3][4] came forward and provided the technological discussions. The goal is to increase the quality and quantity of the crops while optimizing the human labour used. Smart agriculture has proven its superiority over primitive agriculture by reducing cost and increasing efficiency compared to the traditional method. Smart agriculture optimizes the monitoring and measuring of seeds, water, and use of pesticides usage. On the other hand, with improper use of the seed, water, pesticides, and lack of monitoring knowledge, crops are being damaged. Different types of crops need different methods of cultivation [4]. Moreover, the same crop has variations and thus requires different ingredients [1]. Thus, farming demands a decision support system and monitoring system [4][5].

The IoT refers to billions of physical objects connected through the internet for accumulating and utilizing data for decision-making ability [6]. Being ubiquitous by nature, any physical thing can be linked to the internet and operated. Sensor-equipped devices and items are also linked to an IoT platform, that allows collecting data from the devices attached in IoT. Moreover, devices and objects with built-in sensors are connected to an IoT platform, which combines data from various devices and uses intelligence to distribute the most useful information with apps tailored to individual needs.

The complexity inherited in agriculture and a need for a decision support system and monitoring system has been addressed by [7], [8], [9]. To fulfil the aim, in this research a smart agriculture model is developed. Centring the model, this research attempts to investigate the targets that are expected to achieve using smart agriculture. Thus, the question posits, how smart agriculture addresses the vital target of SDGs?

In this research, smart agriculture is defined as an IoT based agriculture system for tracking, monitoring, automating and analyzing the farming land conditions The aim is nothing different from [2][3][4], an attempt to increase the quality and quantity of the crops while optimizing the human labour and utilizing a decision support system for agriculture.

## II. LITERATURE REVIEW

Recent advances in IoT have enabled major strides in agriculture and other agro-based industries. the applications of IoT techniques in agriculture in four categories: controlled environment planting, open-field planting, livestock breeding, and aquaculture and aquaponics [10]. However, Salam et al. advocated IoT as an enabling paradigm for sustainable community development through the interconnection of systems, sensing and communication technologies utilized in IoT [9]. IoT is an envisaged engineered system to utilize natural resources and to protect the natural and environmental systems for a balanced sustainable world. The study [11] presented some practical implementations that suggest IoT shift in agriculture, such as in East Timor cloud-based IoT supports motoring the illegal fishing activity, in Africa animal tracking IoT supports game parks management, in Indonesia IoT enabled digital forestry, In India pumps are interconnected using irrigation IoT for mobile-based irrigation management, Agriculture IoT based on soil moisture sensing to tea crop in Sri-Lanka and Rwanda.

Recently, the study [7] applied an IoT based on low power wireless sensor and the network was based on LoRaWAN protocol. The primary focus was to design a low-cost IoT agriculture application, such as greenhouse sensing and actuation. As the aim was to develop a low-cost IoT based agriculture, the research designed available commercial components and free or open-source software libraries. The purpose was to demonstrate the feasibility of a modular system built with cheap off-the-shelf components. The sensors collected data and stored it in a database located in a cloud service. The collected data can be visualized in real-time by the user with a graphical interface. The reliability of the whole system was proven during a continued experiment with two natural soils, Loamy Sand and Silty Loam. Regarding soil parameters, the system performance has been compared with that of a reference sensor from Sentek.

The applicability of the LoraWAn protocol of [7] was also supported by [8]. The study by [8] has considered agricultural monitoring as an example where IoT has an enabling rolein increasing productivity, efficiency, and output yield. In general, IoT inherits limitations from the power of the devicesin the IoT infrastructure and this study scholarly pointed out the concern in the smart agriculture domain. However, the study applied an experimental comparison is performed between IoT devices with energy harvesting capabilities that use three wireless technologies: IEEE 802.11 g (WiFi 2.4 GHz), IEEE 802.15.4 (Zigbee), and Long Range Wireless Area Network (LoRaWAN), for agricultural monitoring. Four experimentswere conducted to examine the performance of each technology under different environmental conditions. Accordingto the results, LoRaWAN is the optimal wireless technologyto be used in an agricultural monitoring system, when thepower consumption and the network lifetime are a priority.

The study [12] developed a WSN model for watering crops to maximize agriculture by designing and developing a control system that connects the node sensors in the field to data management via smartphone and web application. The model consists of three components,hardware that acts as a control box for connecting to and obtaining agricultural data, a web application that manipulates crop data and field information, and a mobile application that controls crop irrigation. The gathereddata were analyzed using the data mining technique to estimate the appropriate temperature, humidity, and soil moisture for crops in the plan. The result showed that the moisture content in the soil was suitablefor the vegetables, the model reduced costs and increased agricultural productivity.

Gill et. al. proposed a cloud-based autonomic information system for delivering Agriculture-as-a-Service through the use of cloud and big data technologies. Thesystem gathers information from users through pre-configured devices, IoT sensors, and processes it in the cloud usingbig data analytics. The aim is to provide the required information to users automatically. The performance of the proposed systemhas been evaluated in a Cloud environment and experimental results show that the proposed system offers better service and the Quality of Service (QoS) is also better in terms of QoS parameters [13].

The study by [14], aimed at the capabilities of IoT in-field monitoring, such as soil moisture, humidity and temperature of the field. Farmers are expected to take prompt action to manage the field based on the data received from the field. The study used an Arduino Microcontroller board and sensors for soil, temperature and humidity data collection. Data are stored and analyzed for data-driven decision-making in agriculture.

The study by [15] demonstrated the use of IoT in thefield of agriculture. The project used sensors to measurevarious parameters, such as field temperature, humidity, soilmoisture. The data of the farming field is collected andpassed to the server for the store. Based on the data, analysis is done to calculate the water requirement. soil requirementssuch as nitrogen, phosphorous and potassium and fertilizers¿expected advantages, nothing surprising, and optimum cropproduction. Furthermore, based on the intelligence of thesystem also generates the irrigation schedule. The project was implemented in the real world which confirms the enabling role of the IoT in agriculture for improved decision making.

In India, the study [16] aimed to focus on sustainable irrigation due to uncertain climatic conditions. The systemconsists of Raspberry Pi, various sensors, a Pi camera andamotor driver. The purpose of the pi camera captures thevideo and transfers it to the cloud and the soil moisture sensordetects the moisture level and irrigates the various crops ina controlled manner. The data captured using the cameraand the sensor data was the basis the farmers are expected to control the operation.

The study [17] developed a set of agricultural Internet systems with expert guidance. The study purported that the IoT system in agriculture mainly consists of three layers, e.g., perception, transportation, and application. In their research, researchers investigated the key sensors in the perception layer, the application mode of Bluetooth and 4G in the transportation layer, and an intelligent algorithm and an application framework for the application layer. The small-scale farmland experiment is carried out to help farmers in monitoring the plant production process, early warning of main diseases and pests, and rapid diagnosis. [17].

The study [18] explored the possibility of applying IoT foragriculture to trace and track food quality and safety. Mobileapplication for food freshness investigation was successfullydeveloped and the results showed that consumer mobile cameras could be used to test the freshness of food. By applying the IoT technology this information could be shared with allthe consumers and supervisors. [18]

The review of the literature, presented above, suggest that researchers and the IT industry are very keen on revealing a new efficient IoT model for agriculture. This will indeed mature the field. However, the absence or near absence of the discussions on societal and global concerns diminishes the vivid advantages of IoT in agriculture.

## III. RESEARCH METHODOLOGY

In this study, the qualitative technique was used since it is appropriate for problem-centric research. There is a growth in the use of qualitative research techniques in computer studies and the Information Technology discipline, partly because qualitative research offers numerous options to dealing with problem-centric research, such as the experimental method.

The experimental technique was used in this study asa qualitative methodology. To find a useful answer to aproblem, an experiment is carried out. The response to the experiment's query must be included. Various researchers have different points of view on the study. William I.B.Beveridge, for example, defined experimental research as *"a cause of an occurrence to occur under controlled settings in which as many external influences as possible are excluded and close observation is enabled to reveal relationships between phenomena."*

## IV. DESCRIPTION OF TECHNOLOGY AND THINGS OF IoT

In this research, It is feasible to establish a sensor networkthat may be utilized for continuous farm monitoring using IoT devices. Sensors installed on the farm capture data (such as soil moisture, fertilizer, temperature, humidity, acidity, and illumination) and provide real-time data about the land, crop, livestock, and equipment. Farmers can select where their water supplies should be diverted using datacollected from distant sensors. John Deere, the world's largest agricultural equipment company, opted to install GPS sensors in their tractors and other machines in 2001. Sensors mounted on moving machinery can be used to take measurements while in motion. Sensors can be implanted in cattle to monitor stomach acidity and investigate digestive issues.

The smart agriculture model is this research built basedon Wireless Mesh Sensor Network (WMSN). WMSN not only reduced the hard work but also increase the speed.In addition, WMSN is very efficient because it needs a very little amount of power and embedded networking Intelligence [19]. For wireless communication, we used Zigbee 5.8 (ALT5802). It's a compact 5.8 GHz ISM band wireless transceiver module with a transmit power of +30 dBm that supports the entire ZigBee / IEEE
802.15.4 feature [19]. In agriculture, the Implementation of WMSN applications has the potential to boost efficiency, productivity, and profitability while reducing undesired consequences on crops and the changing climate.[20]. The system includes an active Radio-Frequency-Identification (RFID), moisture sensor, temperature sensor [4] [21]. The cultivation area's real-time data will provide an opportunity in making decisions based on an expected average state [20].

The WSN hub is constituted of Intelligent Sensing Devices, a microcontroller, and low-power radio handsets that collect data from the field and send it to a distant destination. In the simulation, we use 4 Sprinklers, 2 Temperature monitors, 2 water level monitors, 2 Motiondetectors and a buzzer. There is a home gateway where all devices are connected through WiFi. Some conditions areset up for activating those devices. The minimum value ofthe water level is set at 5cm. When the water level goesdown to this level the sprinklers will automatically turn on. When the water level reaches 10cm the sprinklers will automatically turn off. The maximum temperature is set upat 35 degrees Celsius. When the temperature goes at 35degrees Celsius the sprinklers will automatically turn on and when the temperature goes down to 25 degrees Celsius the sprinklers are automatically turned off. The first temperaturemonitor and first water level monitor control sprinklers one and two and the second temperature monitor and second water level monitor control sprinklers three and four. When any motion is detected by any of the motion detectors the buzzer is automatically turned on. Following things of IoT were used in the experiment.

*A. WSN (Wireless Sensor Network)*

The wireless sensor network (WSN) is a sensor and networking system that can sense data from the environment and is used for broadcast. The Wireless Sensor Network(WSN) is an infrastructure-free wireless network that uses an ad-hoc deployment of a large range of wireless sensors to monitor system, mechanical, and environmental factors.

WSN uses sensor nodes in conjunction with an integrated CPU to manage and monitor the environment in a specific area. They are linked to the Base Station, which serves as the WSN System's processing unit. A WSN system'sbase station is connected to the Internet to share data. This structure of the device is as follows.

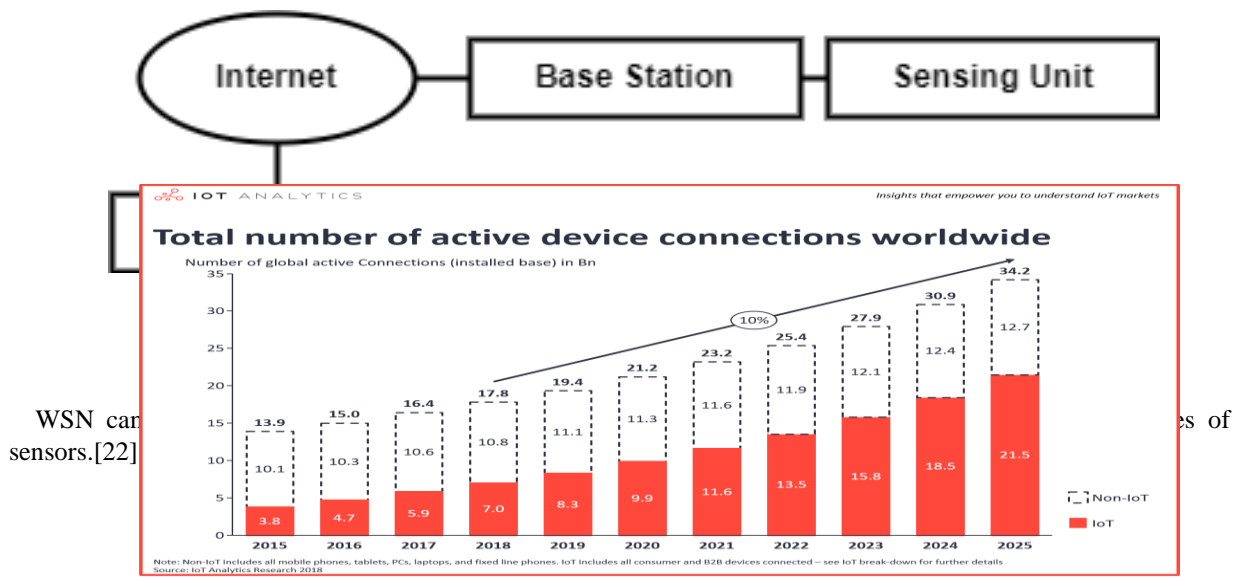

WSN can be used to connect many types of sensors.[22]

Fig2: Benefits of using WSN

According to Figure 2, roughly 75.44 billion devices will be connected to the internet by 2025. In terms of agricultural trends, IoT will be one of them. In IoT-based agriculture applications, items communicate with one another to offer important information from the farm and greenhouse. Some gadgets in these types of systemsperform essential tasks, such as watering. To put it another way, the internet and physical agents have proven to bebeneficial in lowering human factor participation, increasing productivity and lowering expenses.

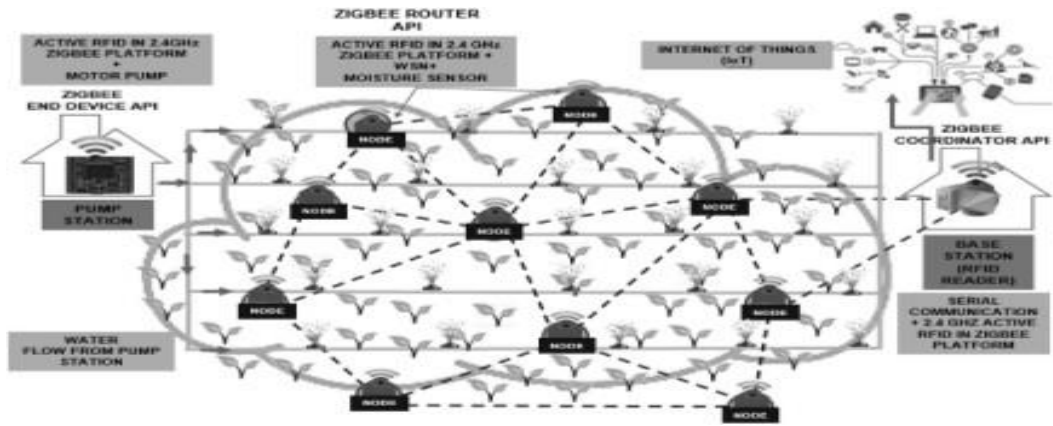

Fig3: Wireless Mesh Network for Agriculture [4]

At the beginning of our study, we simulate our project on tinkercad. Figure 4 represent the initial architecture of our project.

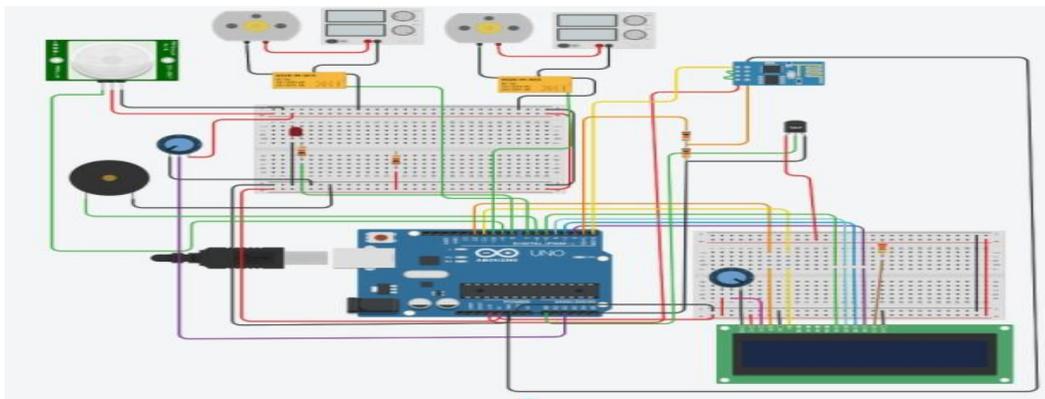

Fig4: Smart Irrigation System.

This is the basic design of our system. The wired connection is shown here but we connect all devices wirelessly. Anothersimulation was done using a cisco packet tracer. Figure 5 is representing that simulation.

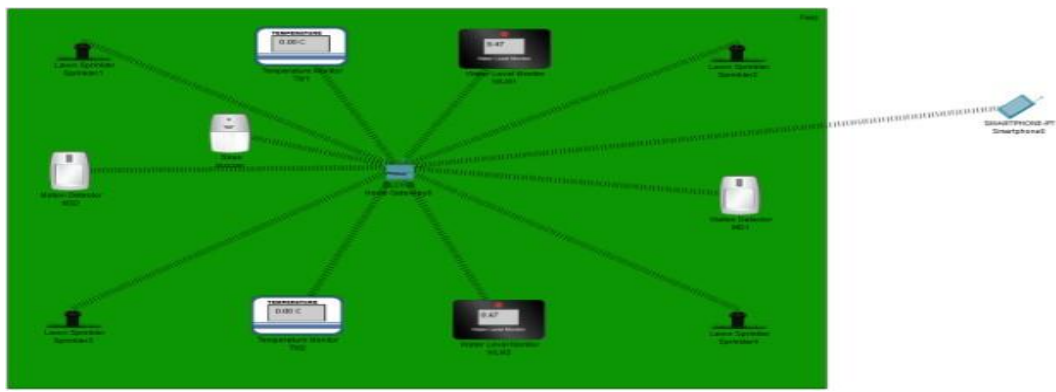

Fig5: Smart Irrigation System.

### B. RFID(Radio Frequency Identification)

RFID is a technology that is used to tag items. RFIDtechnology is made up of tags that generate radio signalsthat RFID devices can detect. RFID can be used to tag WSN nodes in a system, for example. It's not always easyto figure out where a signal is originating from. RFID provides a solution to this problem. RFID is widely used because it is inexpensive, simple to implement, and intuitive to operate.[4].

### C. Zigbee 5.8

Zigbee is a very well recognized mesh networking standard for connecting sensors, instruments and control systems. It has a packet-based protocol designed to provide a simple architecture for a reliable, secure and low power wireless networks.[33]

The importance of Zigbee lies in its ability to provide a low data rate, low battery consumption at low cost. Its main application is in the field of home automation, industry and remote control, medical assistance and other wireless sensor applications as it is mainly used wireless sensor and wireless personal area network applications (WPAN)[34]

To achieve Sustainable Agricultural Development, it is very important to implement automation of agricultural data collection through sensor networks. As mentioned, Zigbee technology is capable and is a very suitable implementation architecture for such data collection.[35]

It's a 5.8 GHz ISM band wireless digital information and communication technology with a transmit power of +30 dBm that supports the whole ZigBee / IEEE 802.15.4 featureset. It's built for OEMs that want to add a strong 5.8 GHz wireless connectivity to their device fast and inexpensively, utilizing comprehensive networking features for IoT platforms like the Internet of Things devices.[4][24][25].

### D. Logic design

Implemented code in the devices

```
//Initialization:
if(mois<0.5)
  {
   digitalWrite(motor,HIGH);
   lcd.setCursor(0,1);
   lcd.print("low mois, Motor on");
  }
 else if(temp>=75)
  {
   digitalWrite(sprinkler,HIGH);
   digitalWrite(led,HIGH);
  }

else
   {
    digitalWrite(sprinkler,LOW);
    digitalWrite(led,LOW);
    digitalWrite(motor,LOW);
   }
   inf= digitalRead(sensor);
   if(inf==1)
   {digitalWrite(buzz,HIGH);}
   else if(inf==0)
   {digitalWrite(buzz,LOW);}

 //Temperature:
 lcd.createChar(0,degree);
 lcd.clear();
 lcd.print("Temp:");
 lcd.print(temp);
 lcd.write(byte(0));
 lcd.print("C");
 if(temp>=75)
 {lcd.setCursor(0,1);
 lcd.print("FIRE !EVACUATE!!");}

 //Moisture
 lcd.clear();
 lcd.print("Moisture:");
 lcd.print(mois);
 lcd.write(byte(0));
 lcd.print("%");
```

```
   if(mois<0.5)
   {lcd.setCursor(0,1);
lcd.print("low moisture, Motor on");
   }

//Air Quality:
   lcd.clear();
   lcd.print("AirQ:");
   lcd.print(air);
   lcd.print("ppm");
   lcd.setCursor(0,1);
   //Door
    if(inf==1)
   {
        lcd.print("Intruder");
```

Logic Pseudocode:

*Step 1:* Initialization of all sensors with collected sensor's value.
*Step 2:* If humidity is less than .5 motor will on.
*Step 3: Else*-If temperature is greater than 75 sprinklers will on.
*Step 4:* Else motor and sprinkler both remain off.
*Step 5:* If any objects come around inf (Passive infrared sensor) == 1 a buzzer will make noise
*Step 6:* Else buzzer remains off.
*Step 7:* Go to step 1 after every 30 minutes.

## V. RESULTS

The system begins with detecting the field's state by the sensors and reports the outcome to the base to cite-jawad2017energy. Based on these parameter values, first, the system will compare those values with predefined set standard values and the system will start or stop functioning based on some conditions. Like the minimum value of the water level is set at 5cm. When the water level goes down to this level the sprinklers will automatically turn on. When the water level reaches 10cm the sprinklers will automatically turn off. The maximum temperature is set up at 35 degree

Celsius. When the temperature goes at 35 degrees Celsius the sprinklers will automatically turn on and when the temperature goes down to 25 degrees Celsius the sprinklers are automatically turned off. There are other conditions for providing pesticides also. In our research, there is a shield for our field. An ultrasonic sensor is implemented. An ultrasonic sensor is a device that uses ultrasonic sound waves to determine the distance to an item. A transducer is used in an ultrasonic sensor to emit and receive ultrasonic pulses that communicate information about the proximity of an item. If any animal tries to enter the field's certain area the ultrasonic sensor will send a signal to a buzzer and the buzzer will make a loud sound to scare the animal make it get out. [26].

In the developed project, we have tried to measure all the constraints that are required to be handled to build a Smart Agriculture System to meet SDGs. In the sensor model we designed, we have measured the moisture level, temperature, developed an alarm mechanism to keep small animals away, as well handle severe temperature fluctuations as in case of a fire with the motor triggering sprinkler system to extinguish it. We have performed simulations for different scenarios and explained how each is handled as in figures 8 – 12 below. The simulation has been run as real-time monitoring and the findings of the simulation are also stored in the server, just how real data would be backed up. We have set up the model to take temperature and humidity measurements take every 30 minutes for a total of 24 hours. As you can see in figure 8, the temperature detected is shown in degrees Celsius.

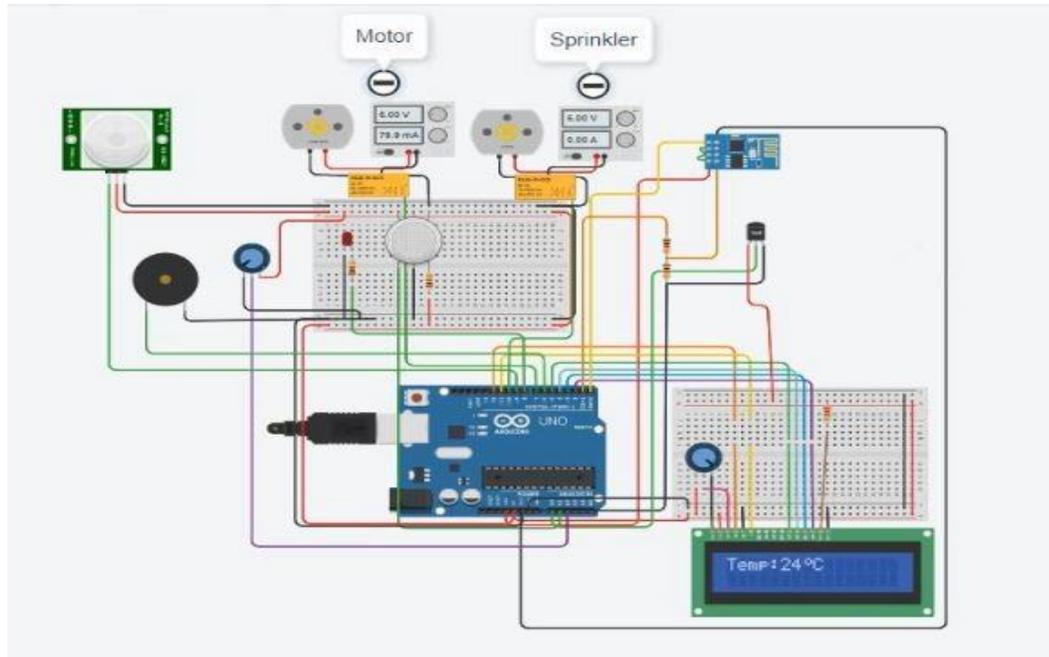

Fig8: Temperature monitoring

In the following figure 9, we have displayed the humidity reading. One of the main reasons to take humidity into account is because of SGD **2.4 Ensure sustainable food production systems**. Humidity is a vital factor to provide good agricultural yield and affects how long the production yield can be sustained properly. humidity below 50% can be devastating for crop growth over an extended period [5], so we have prioritized it in our model. As we can see from the reading at 2:05 pm om the figure below, the humidity is at its lowest (61.32%) when the temperature is high (37.48 °C).

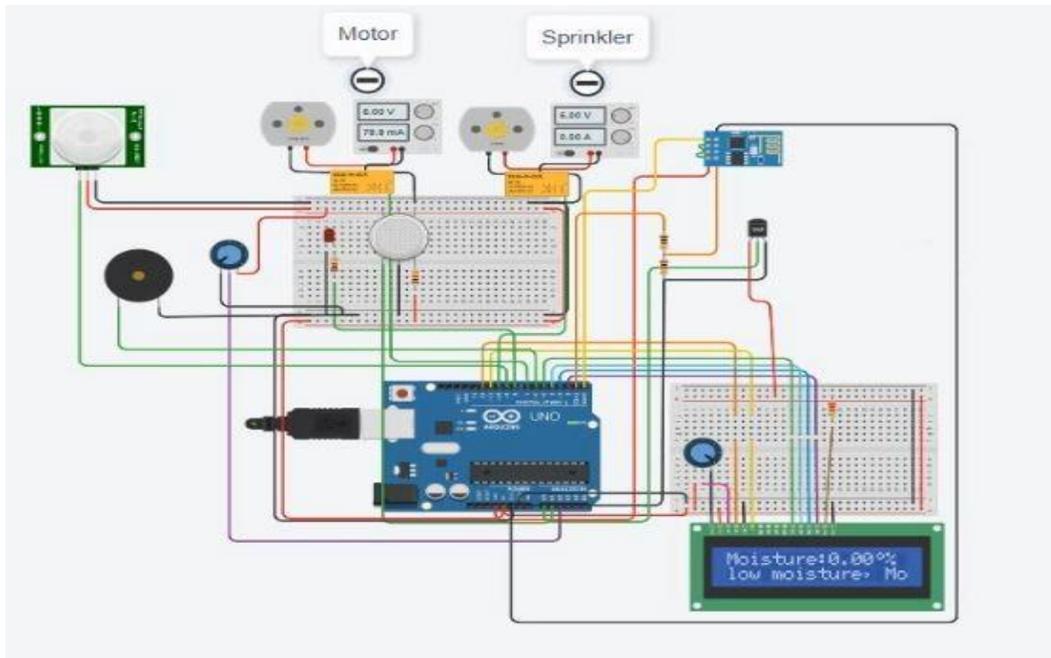

Fig9: Humidity monitoring

Figure 10 shows in case of extreme temperature changes across the machinery (like a spike to a 75 ºC due to a fire), the system will automatically start spreading water via a sprinkler system to ensure overheating of the machinery save them from burning. The smoke alarm that will be triggered is also visible in the diagram below. This same mechanism will also be triggered if the temperature nearby triggers (like a fire across the field), which we have planned will come to aid in extinguishing small fires across the fields as long as they are within the sensors detection range. This aims to achieve SDG **9.4 - Greater adaptation of clean and environmentally sound technologies and industrial** processes because this system is very environmentally friendly and also is scalable to fit mechanisms for large industrial areas.

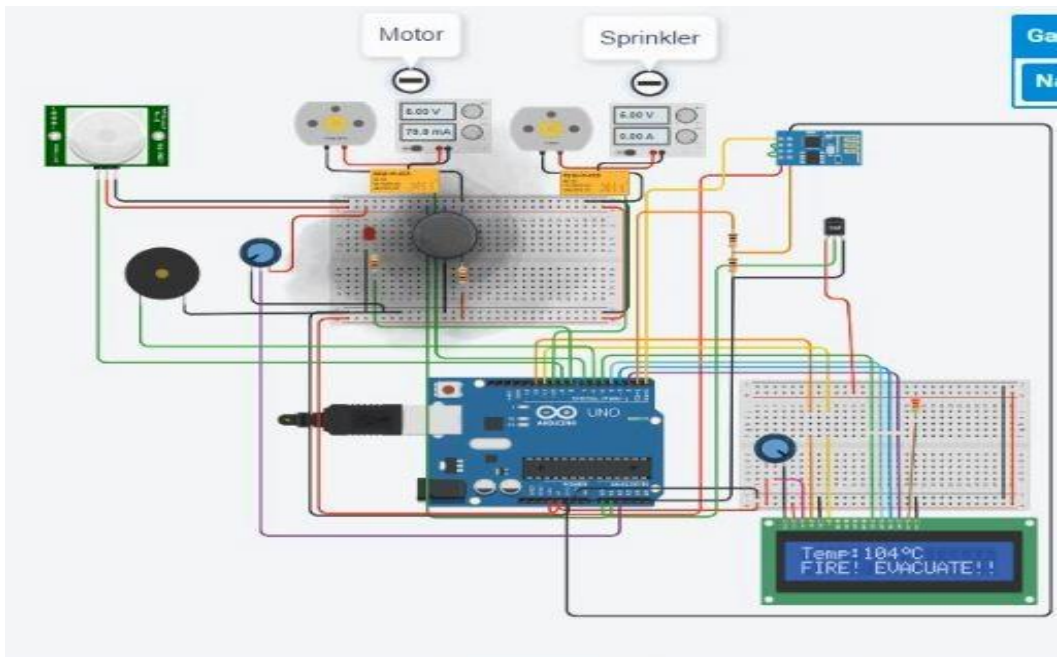

Fig10: High-Temperature monitoring

We have shown in Figure 11 below what will happen if the moisture level is too high for a certain period (the opposite has also been considered in terms of moisture level and humidity). The main goal is to ensure that water supply is always consistent so that neither the crops starve of water for long periods, as well as that wastage of water is not done by properly supplying the water until moisture levels come to an acceptable limit. Our main goals with monitoring humidity and ensuring proper supply of water are done to the crops (on time, low/no wastage) is to achieve the SDGs **End of poverty with the use of the latest technology and natural resources (1.2), Ensure sustainable food production systems (2.4)** and **Sustainable management and efficient use of natural resources (12.2).**

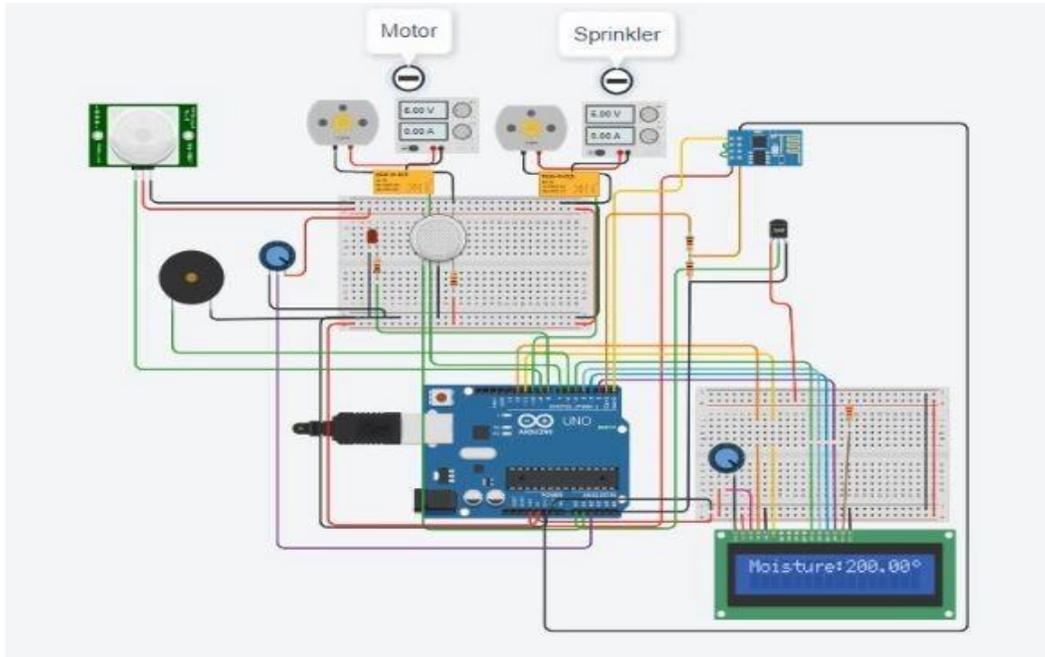

Fig11: High Humidity monitoring

Apart from supplying the water correctly to the crops and also ensuring proper temperature control, it is also a very important step to ensure keeping the crops safe from small animals/pests, and also try to do so without the use of pesticides, because we want to use an environmentally clean system that can be industrially scaled (SDGs **9.4 & 9.5**). We have thought of a solution to do so using a motion trigger system with an accompanied buzzer. Figure 12 below shows a subsystem of our architecture that contains a motion detector. Whenever an animal of a certain size tries to go near the crops, it will be detected and the buzzer will make a sound to scare the animal away. Currently, this is only using a loud sound but can be a modifier easily to use an ultrasonic range of sound or vibrations to scare the animals away in more advanced models in future.

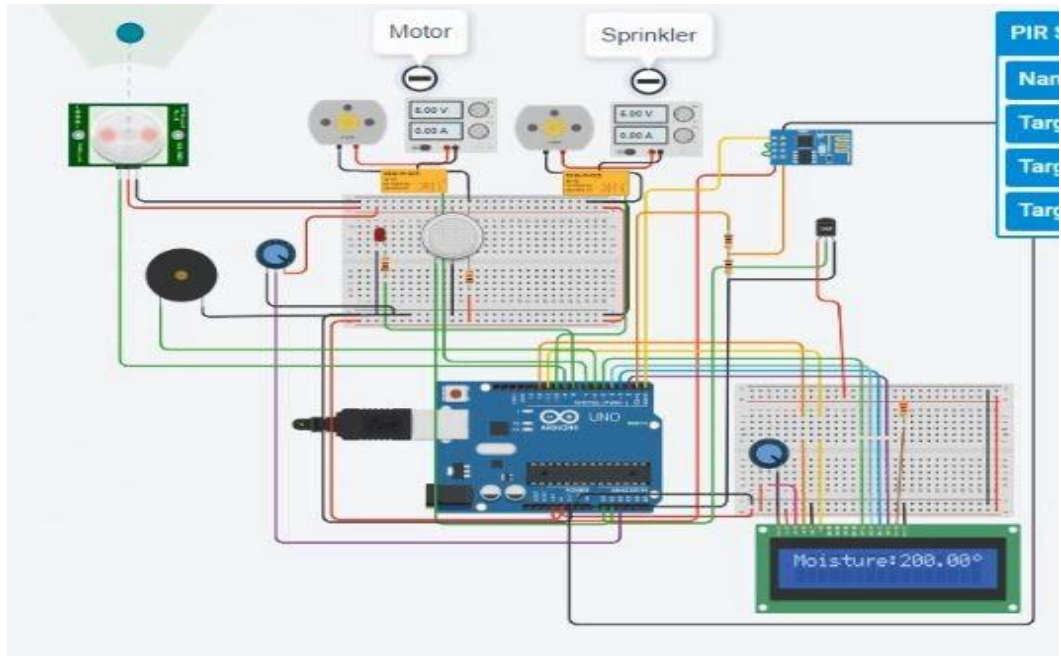

Fig12: Animal monitoring

## VI. DISCUSSION

The discussions of this study are of two folds; first, the discussion concentrates on the applicability of the smart agriculturemode in the field. In the model 4 Sprinklers, 2 Temperature monitors, 2 water level monitors, 2 Motion detectors and a buzzer were used. Temperature, high-temperature, humidity,high-humidity and animal detection experiment. The model aligns with the studies [4][24][25].

The entire system will be monitored by IoT using WSN and RFID technologies. Because there is no unique identification of WSN nodes by default, RFID technology was utilized to identify each one. Because of solar light variations, soil structure differences, and a variety of other factors, the moisture and temperature levels of the agriculture field may vary in different locations. As a result, it's critical to identify each WSN node separately since various nodes may require different amounts of water, seeds, and pesticides [21][25]. The proposed system communicates with the hardware and software and automatically sends information to the Base station. The moisture and temperature data from the field will be detected by WSN nodes and sent to the farm through a wireless network. The suggested systems employ a highly intelligent paradigm in which nodes are constantly in the standby state. Thisis not only incredibly energy efficient for the system, butit also ensures that the battery usage is low. If the sensor detects dry soil, it will immediately activate and gather data before sending it to the station. It detects the node aftertransmitting the data to the base station, allocates whichnode sent the data, and begins the necessary operation.[27]. Figure 3 show the architecture for the wireless mesh network.

The second point of discussion directs to SDG targetsthat are expected to be achieved using the proposed model in this study. From our observation, The outcome of the experimental model we presented in this research has made the following Sustainable Development Goals the main area of interest in this study:

- Partial Contribution to Goal 1 (No poverty) and Goal 2(Zero hunger)
- Goal 8 (work and economic growth)
- Goal 9 (Industry, Innovation and Infrastructure)
- Goal 12 (Responsible Consumption and Production)

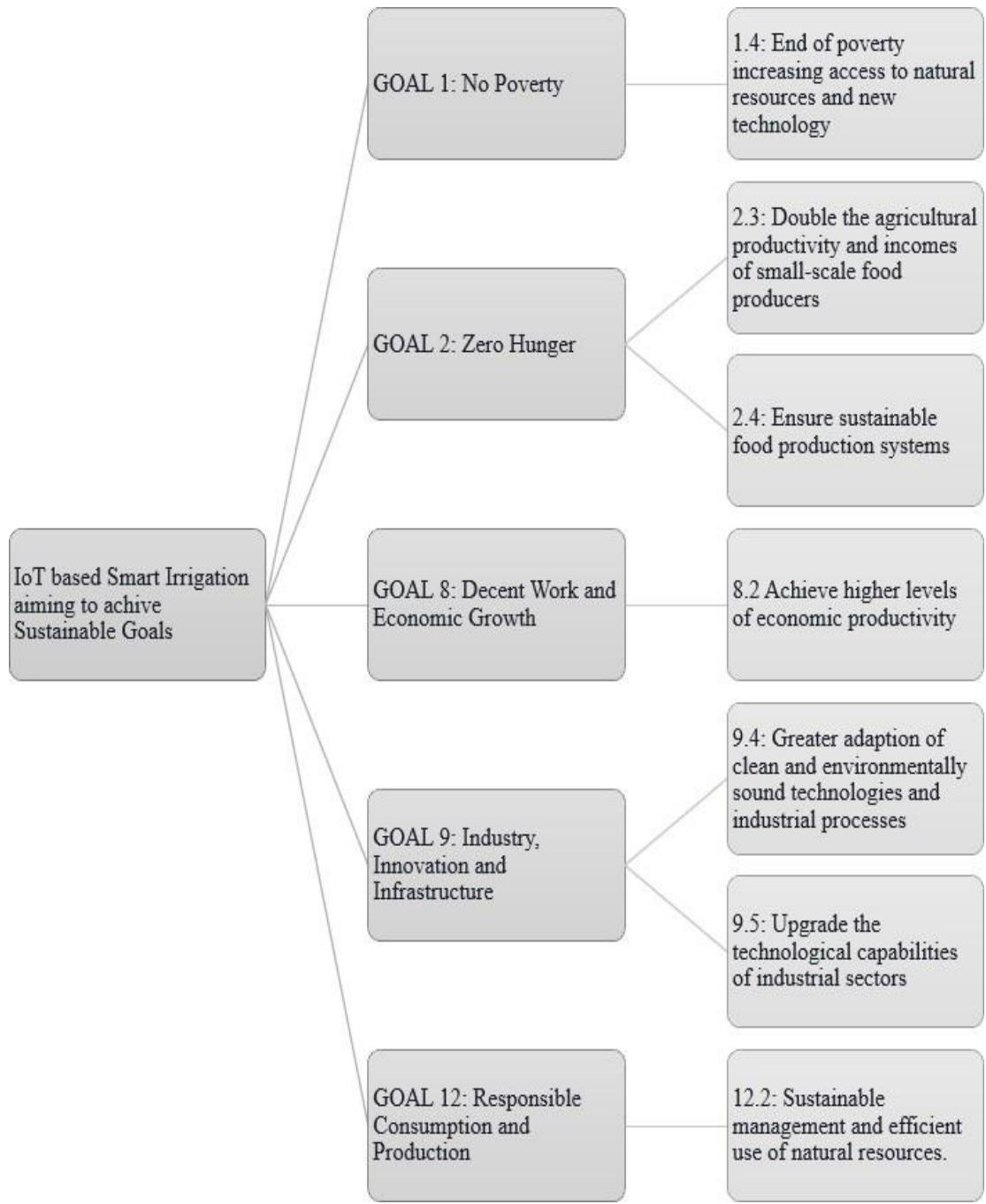

Fig 9: Research focus on SDGs

After an investigation of the experimental results, we can discuss how the outcomes of our experimental model have an impact and act to achieve SDGs. Our model aims to:

1) **Goal 1.4 - Bring an end to poverty with the use of the latest technology and natural resources:** The IoT controlled smart agricultural monitoring system primarily aims to bring together the usage of the latest wireless networking technology together with IoT to improve and implement a low-cost monitoring system. The sensors are also designed to make proper use of natural resources to detect changes in and improve food production. This will contribute to the goal of bringing an end to poverty by spreading the usage of technology in low economic communities and more efficient usage of natural resources.
2) **Goal 2.3 - Increase the agricultural yield and income for farmers starting from the smallest level:** Our model aims to increase the productivity of the field yields and result in an increase in income for the farmers at the smallest level. We have tried to focus the model to make use of natural resources and build a strong and smart irrigation system, something which the small farmers cannot typically achieve because of the lack of funds and manpower.
3) **Goal 2.4 - Assurance of sustainable crop yield:** Since the model is inexpensive for maintenance and works mostly with natural resources, it is sustainable. Our model aims to help the agricultural ecosystem and small farms to be stronger in terms of adaptability to climate, drought, and moisture issues that impact and affect the land and food quality. Moreover, the model has much room to improve with the collection of statistical data and so making the future iterations more sustainable.
4) **Goal 8.2 - Achieve higher levels of productivity:** Our model aims to contribute to the goal of achieving higher levels of economic productivity using introduction and increasing the reach for technological growth, as well as increasing the levels of innovation in the smaller community using usage of IoT.
5) **Goal 9.4 - Adapting to clean and environmentally sound technologies:** The smart agriculture model makes use of natural resources and environmental factors to perform irrigation as well crop protection. And it has the option to industrially expand this over a very large network, thus making it scalable and easy to adapt across multiple industrial processes.
6) **Goal 9.5 - Upgrading technological capabilities across industrial sectors:** The IoT based model presented in this research has the option to contribute to upgrading technological capabilities across variations of the wireless sensor networks used as well smart agricultural methods. We have used moisture sensors along with temperature sensors, to both control motors and trigger proximity alarms. These can be expanded for various industrial sectors and increase the technological capabilities for different use cases.
7) **Goal 12.2 - Sustainable management of natural resources:** The model aims to make proper usage of natural resources available, and perform most of the smart automation via them. It aims to minimize the wastage of water and make the increased usage of climate and temperature available to us, and this makes the consumption of natural resources minimal and long term usage sustainable.

Apart from the direct contributions to SDGs, there are some other areas the model upholds. A poor harvest occurs due to a lack of water in a specific section of the agriculture field or over flooding of water because farmers do not know when and how much water to give in their field. To address this issue, WMSN and active RFID have been implemented. We presented a real-time autonomous system to monitor the field and regulate the irrigation system in this study. This system will be capable of replacing human-to-human, machine-to-machine, and human-to-machine architecture. We will be able to recognize which areas of the land require watering with the help of active RFID, and WMSN combines the reliability of hardwiring with the versatility of wireless networking without sacrificing speed. At the same time, WMSN is both cost-effective and battery-powered. Farmers will be able to see real-time data on the status of their fields, which will aid them in making decisions about when to irrigate, allowing them to save water and avoid overwatering.

## VI. CONTRIBUTIONS

This study offers several contributions; those are listed below:

### A. Practical and national implications

The expected primary practical benefit is obviously for the stakeholders related to agriculture, the farmers and the government. The next practical benefit for the IT practitioners, the experimental model is expected to provide insights into how cost-effective smart agriculture can be developed. Lastly, cities with having lass land for cultivation can apply the concept of smart agriculture to increase productivity, monitor fields and control the use of insecticides.

### B. Theoretical contribution

[9] pointed out that there are big gaps in our understanding of enabling role of smart agriculture in the area of sustainability, climate change and natural resource utilization. This research, taking into SDG as a theoretical base, attempts to provide light in the path of SDG achievement.

The study, on the other hand, focuses on the development of mid-range theories. Gregor defines midrange theory as a moderately abstract theory, has a narrow scope, and may readily lead to testable hypotheses [28]. Midrange theory is essential because it deals with practice-based dis- ziplines, notably in the social sciences [29] [28]. The mid-range theory has been noted as important as midrange theory deals with practice-based disciplines, especially in a social sciencescenario [29], and it is also noted that social phenomenon consists of social relations and practices that are deeply rooted in technology. Mobile app as a branch of technologies also relates to business Process Management of the organizations and social practices in each context.

### C. Methodological contributions

This research provides a thorough literature review of smart agriculture related to understanding the current knowledge gap. Barki et al. emphasize that researchers should focus on methodological advancement as there were insufficient guidelines for construct creation in the research studies [30]. Following these ideas, this research provides a well-documented research journey outlining the choice of base theory the social capital theory, the justifications of the qualitative methodology and an experimental approach.

## VII. CONCLUSION

This research presents a model of IoT enabled smart agriculture. centring the model, we provide a descriptive analysis of how the IoT based smart agriculture addresses the SDG targets. The analysis suggests goal 1-No no poverty, goal 2-Zero hunger, goal 8-work and economic growth, goal 9-Industry, Innovation and Infrastructure and lastly goal -12 Responsible Consumption and Production is some of the SDG targets that are expected to be addressed by the model.

Like much other research, this research also inherits some limitations. This research is a simulation-based experiment; thus one might criticize that the study lacks empirical knowledge. However, Robinson et. al. purported that in the scenario where the problem description rarely contains sufficient information to make key decisions about the level of granularity, the conceptual model provides the opportunity to provide insightful discussions, that is appropriate for specifying behaviour characteristics [31]. In this research, the problem formulates that the notion of SDG achievement was not extended in smart agriculture research discussions. In this research, the conceptual model presented, provides an opportunity to discuss the practicality of IoT in achieving SDG targets. In this study, the main benefit of the conceptual model is to have all available IoT devices integrated into an agriculture scenario. The interpretations of this study at hand are evidence for decision-making. The conceptual smart agriculture model becomes central to managing activities in the farming of different phases [32]. The findings, especially other limitations the SDG goals, expected to achieve in the study is not tested, neither qualitatively nor quantitatively, however, the findings are observational.

Future works and goals are therefore put smart agriculture in a real-world setting. Addressed SDG targets should be evaluated using qualitative or quantitative criteria. Climate-related negative and positive aspects should be investigated utilizing IoT-based sensors. To achieve the SDGs, artificial intelligence and IoT might be combined. The limitations of IoT devices and connectivity in smart agriculture should potentially be investigated further.

However, following visible successes of IoT enable smart agriculture as addressed by Salam et al. [9] we address the need for SDG achievement utilizing smart agriculture by providing a conceptual model of smart agriculture. To our knowledge, this is the first study to look at how IoT-based smart agriculture may help achieve the SDGs. The findings will be useful to both developed and developing countries in the future for a more sustainable planet.